\documentclass[a4paper, twocolumn]{article}
\usepackage[utf8]{inputenc}
\usepackage{cite}
\usepackage{graphicx}
\usepackage{amssymb}
\usepackage{amsfonts}
\usepackage{amsmath}
\usepackage{bm}
\usepackage{longtable}
\usepackage{epsfig}
\usepackage{multirow}
\usepackage{authblk}%
\usepackage[paperwidth=19.5cm,paperheight=27.0cm,top=2.0cm,bottom=2.0cm,left=1.5cm,right=1.5cm]{geometry}
\usepackage[switch]{lineno}
\usepackage{fancyhdr}
\begin{document}
%\begin{CJK*}{UTF8}{song}

\title{Derivation of the Schr\"{o}dinger equation based on a fluidic continuum model of vacuum and a sink model of particles }
\author{Xiao-Song Wang}
\setcounter{footnote}{0}
\affil{{\normalsize Institute of Mechanical and Power Engineering, Henan Polytechnic University, Jiaozuo, Henan Province, 454000, China}}%
%\date{}
\date{Apr. 14th, 2024}
%\maketitle

%%%%%%%%%%%%% line numbering
%\linenumbers
%%%%%%%%%%%%% line numbering
\twocolumn [
	\begin{@twocolumnfalse}
		\maketitle
		\begin{abstract}
			\newgeometry{left=1.0cm, right=1.0cm}%
We propose a fluidic continuum model of vacuum and a sink flow model of microscopic particles. The movements of a microscopic particle driven by a stochastic force was studied based on stochastic mechanics. We show that there exists a generalized Schr\"{o}dinger equation for the microscopic particle.

\

keywords: Schr\"{o}dinger equation; stochastic mechanics; Hamilton-Jacobi equation; Langevin equation; sink; ether; Planck constant.
	  \end{abstract}
	\end{@twocolumnfalse}
%abstract
%\vspace*{6pt}

\

%\noindent
%PACS:

%\thanks{$*$ E-mail: }

\

\

]

%%%%%%%%%%%%%%%%%%%%%%%%%%%%%%%%%%%%%%%%%%%%%%%%%%%%%%%%%%%%%%%%%%%%%%%%%%
\section{Introduction}

\newtheorem{assumption_my}{\bfseries Assumption}

\newtheorem{definition_my}[assumption_my]{\bfseries Definition}

\newtheorem{lemma_my}[assumption_my]{\bfseries Lemma}

\newtheorem{proposition_my}[assumption_my]{\bfseries Proposition}

\newtheorem{theorem_my}[assumption_my]{\bfseries Theorem}

\newtheorem{wcorollary_my}[assumption_my]{\bfseries Corollary}

The Schr\"{o}dinger equation for a non-relativistic particle moving
in a potential $V(\mbox{\upshape\bfseries{x}})$ can be written as \cite{Landau-Lifshitz1958}
\begin{equation}\label{Schrodinger equation 10-10}
i \hbar \frac{\partial \psi}{\partial t} =
-\frac{\hbar^2}{2m}\nabla^2\psi + V(\mbox{\upshape\bfseries{x}})\psi,
\end{equation}
where $t$ is time, $\mbox{\upshape\bfseries{x}}$ is a point in space, $\psi(\mbox{\upshape\bfseries{x}},t)$ is the wave function, $m$ is the mass of the particle, $V(\mbox{\upshape\bfseries{x}})$ is the potential, $h$ is the Planck constant, $\hbar = h/2\pi$
 and $\nabla^2 = \partial^2 / \partial x^2 + \partial^2 / \partial y^2 + \partial^2 / \partial z^2$
is the Laplace operator.

In a remarkable paper \cite{Nelson1966}, E. Nelson derived the Schr\"{o}dinger equation
Eq.(\ref{Schrodinger equation 10-10}) based on two main hypotheses \cite{Nelson1966}.
The first hypothesis is that the position of a Brownian particle satisfies the Smoluchowski equation.
The second hypothesis is that the diffusion coefficient $\nu$ of the Wiener process
can be written as $\nu = \hbar/2m$, where $m$ is the mass of the particle, $\hbar = h/2\pi$, $h$ is the
Planck constant. Nelson's stochastic mechanics \cite{Nelson1966,Nelson1972,Nelson1985} was further developed
\cite{NelsonE2012}.

In previous works of stochastic mechanics \cite{NelsonE2012}, these two hypotheses can not be explained.
Is it possible for us to derive Nelson's two hypotheses based on some assumptions about the nonempty vacuum? The purpose of this paper is to propose a derivation of Nelson's hypotheses. Further, considering the mass-increasing effects, we show that there exists a nonlinear Schr\"{o}dinger equation for microscopic particles. As a byproduct, the Planck constant $h$ is calculated theoretically.

\section{A fluidic continuum model of vacuum and a sink flow model of microscopic particles \label{vacuum 20}}
Many philosophers and scientists, such as Laozi \cite{Laozi1995}, Thales,
Anaximenes, etc., believed that everything in the universe is made of a
kind of fundamental substance \cite{WhittakerE1951}. Descartes was the
first to bring the concept of ether into science by suggesting that
it has mechanical properties \cite{WhittakerE1951}. Descartes
interpreted the celestial motions of celestial bodies based on the
hypothesis that the universe is filled by an fluidic vortex ether.
  After Newton's law of gravitation was published in
1687 \cite{Newton-a}, this action-at-a-distance theory was criticized
by the French Cartesians \cite{WhittakerE1951}. Newton admitted that
his law did not touch on the mechanism of
gravitation \cite{Cohen1980}. He tried to obtain a derivation of his
law based on Descartes' scientific research program. At last, he
proved that Descartes' vortex ether hypothesis could not explain
celestial motions properly \cite{Newton-a}. Newton himself even
suggested an explanation of gravity based on the action of an
ether pervading the space \cite{Newton-b,Cohen1980}. Euler
attempted to explain gravity based on some hypotheses of a fluidic
ether  \cite{WhittakerE1951}.

Since quantum theory shows that the vacuum is not empty and has physical
effects, e.g., the Casimir effect, it is valuable to probe the vacuum by introducing the following
assumption \cite{WangXS200810}.
\begin{assumption_my}\label{assumption Omega0 50-10}
Suppose the universe is filled by a fluidic substratum.
\end{assumption_my}

This fluidic substratum may be named the $\Omega(0)$ substratum in order to
distinguish it from the Cartesian ether.

Suppose that a velocity field of a fluid is continuous and finite at
all points of the space, with the exception of individual isolated
points. Then these isolated points are called singularities in this fluid. Suppose there exists
a singularity at point $P_0=(x_0,y_0,z_0)$.
 If the velocity field of the singularity at point $P=(x,y,z)$ is
\begin{math}
\mbox{\upshape\bfseries{u}}(x,y,z,t)=(Q/4\pi
r^2)\hat{\mbox{\upshape\bfseries{r}}},
\end{math}
where $r=\sqrt{(x-x_0)^2+(y-y_0)^2+(z-z_0)^2}$,
$\hat{\mbox{\upshape\bfseries{r}}}$
 denotes the unit vector directed outward along the line
from the singularity to the point $P=(x,y,z)$, then we call this
singularity a sink if $Q<0$. $Q$ is called the strength of the sink.

Further, we introduce the following assumption  \cite{WangXS200810}.
\begin{assumption_my}\label{assumption monad 50-10}
All the microscopic particles were made up of a kind of elementary
sinks in the $\Omega(0)$ substratum. These elementary sinks were created simultaneously. The
initial masses and the strengths of the elementary sinks are the same.
\end{assumption_my}

We may call these elementary sinks as monads.
Suppose that a particle with mass $m$ is composed of $N$ monads. We have the following relationships \cite{WangXS200810}:
\begin{equation}
\frac{dm}{dt} = \frac{\rho_{0} q_{0}}{m_{0}(t)} m(t),\label{monad 20-30}
\end{equation}
where $m(t)$ is the mass of a particle at time
$t$, $m_0(t)$ is the mass of monad at time $t$, $-q_0( q_0 > 0)$ is
the strength of a monad, $\rho_{0}$ is the density of the substratum, $t\geqslant 0$.

If Assumption \ref{assumption monad 50-10} is valid, then, the equation of motion of a particle is \cite{WangXS200810}
\begin{equation}\label{motion 20-10}
m(t)\frac{d \mathbf{v}_{p}}{d t}
= - \frac{ \rho_{0} q_{0}}{m_{0}(t)} m(t) \mathbf{v}_{p} + \mathbf{F},
\end{equation}
where $m_0(t)$ is the mass of monad at time $t$, $-q_0$ is the
strength of a monad, $m(t)$ is the mass of the particle at time $t$,
$\mathbf{v}_{p}$ is the velocity of the particle,
$\mathbf{F}$ denotes other forces.

From Eq.(\ref{motion 20-10}), we see that there exists a universal damping force
\begin{equation}\label{universal damping force}
\mathbf{F}_d = -\frac{\rho_{0}q_0}{m_0}m\mathbf{v}_{p}
\end{equation}
 exerted on each particle by the $\Omega(0)$ substratum.

\section{The Langevin equation model of a Brownian particle moving in an external force field}
In this section, we study the stochastic Newtonian mechanics of a particle in a potential by a method similar to Nelson's stochastic mechanics \cite{Nelson1966,Nelson1972,Nelson1985}.

Suppose that a microscopic particle is moving in an external force field $\mathbf{F}(\mathbf{x},t)$ in
the $\Omega(0)$ substratum. In order to describe the motion of the microscopic
particle, let us introduce a Cartesian coordinate system $\{ o, x_1, x_2, x_3 \}$ which is attached to the static substratum at infinity.
 Let $\mathbf{x}(t)$ denote the position of the Brownian
particle at time $t$. We assume that the velocity
$\mathbf{v}_{p}=d\mathbf{x}/dt$
exists. Suppose there are also a damping force
$\mathbf{F}_{2}$ and a random force
$\boldsymbol{\xi}(t)$ exerting on the particle. Then, according to Eq.(\ref{motion 20-10}),
the motion of the particle can be described by the Langevin equation \cite{Chandrasekhar1943}
\begin{equation}\label{Langevin equation 50-10}
m\frac{d^2\mathbf{x}}{dt^2}  = - \frac{ \rho_{0} q_{0}}{m_{0}(t)} m(t) \mathbf{v}_{p} +
\mathbf{F}_{2} + \mathbf{F}(\mathbf{x},t) + \boldsymbol{\xi}(t),
\end{equation}
where $m$ is the mass of the particle.

We assume that the force
$\mathbf{F}(\mathbf{x},t)$  is a continuous function of
$\mathbf{x}$ and $t$.
Suppose that the damping force $\mathbf{F}_{2}$
exerted on the microscopic particle by
\begin{equation}\label{damping force 50-10}
\mathbf{F}_{2} = - f_{2}m \mathbf{v}_{p},
\end{equation}
where $f_{2} \geq 0$ is a constant.

Using Eq.(\ref{damping force 50-10}), Eq.(\ref{Langevin equation 50-10}) can be written as
\begin{equation}\label{Langevin equation 50-40}
m\frac{d^2\mathbf{x}}{dt^2}  = - f \mathbf{v} +
\mathbf{F}(\mathbf{x},t) +\boldsymbol{\xi}(t),
\end{equation}
where \begin{math}f = \left ( \rho_{0} q_{0}/m_{0} + f_{2} \right ) m.\end{math}

Next, we need a proper mathematical model of the rapidly fluctuating
and highly irregular force $\boldsymbol{\xi}(t)$ to
ensure that Eq.(\ref{Langevin equation 50-10}) is mathematically
explicit. Inspired by the Ornstein-Uhlenbeck theory  \cite{UhlenbeckOrnstein1930,Kallenberg1997} of microscopic motion, it is natural to assume that the random force
$\boldsymbol{\xi}(t)$ exerted on a microscopic particle
by the substratum $\Omega(0)$ has similar properties as the random force exerted on a
microscopic particle immersed in a classical fluid by the fluid.
Thus, we make the following assumptions.
\begin{assumption_my}\label{random force assumption 50-10}
We assume that the random force $\boldsymbol{\xi}(t)$
exerted on the particle by the substratum $\Omega(0)$ is a three-dimensional Gaussian
white noise  \cite{Gelfand-Vilenkin1961,Soong1973,Arnold1974,Gardiner2004} and the strength $\eta_{i}^{2}$, $\eta_{i}>0$, $i=1, 2, 3,$ of the $i$th component of $\boldsymbol{\xi}(t)$ is
\begin{equation}\label{strength assumption 50-10}
  \eta_{i}^{2} = 2 f k_{0} T_{0}, \quad i=1, 2, 3,
\end{equation}
where $f$ is the damping coefficient of the damping force exerted on the
particle by the medium, $k_{0}$ is a constant similar
to the Boltzmann constant $k_{B}$ which depends on the particles which constitute the substratum $\Omega(0)$, $T_{0}$
is the equilibrium temperature of the substratum $\Omega(0)$.
\end{assumption_my}

For convenience, we introduce notations as
$\sigma_{1}=\eta_{i}$, $i=1, 2, 3$.
 According to Assumption \ref{random force assumption 50-10}, the correlation
function $R_{i}(t,s)$ of the $i$th component of $\boldsymbol{\xi}(t)$ is
\begin{equation}\label{correlation function of Gaussian white noise 3-10}
R_{i}(t,s) = E[\boldsymbol{\xi}_{i}(t)\boldsymbol{\xi}_{i}(s)] = \sigma_{1}^{2}\delta(t-s),
\end{equation}
where $\delta(t)$ is the Dirac delta function, $\boldsymbol{\xi}_{i}(t)$ is $i$th component of $\boldsymbol{\xi}(t)$.

For convenience, we introduce the following notations
\begin{equation}\label{diffusion coefficient 3-20}
\nu_{1} = \frac{\sigma_{1}^{2}}{2} = f k_{0} T_{0}.
\end{equation}

The three-dimensional Gaussian white noise $\boldsymbol{\xi}(t)$ is the generalized derivative
of a Wiener process $\mathbf{N}(t)$ \cite{Gelfand-Vilenkin1961,Arnold1974}. We can write
formally \cite{Gelfand-Vilenkin1961,Soong1973,Arnold1974,Gardiner2004}
\begin{equation}\label{random force 3-20}
\boldsymbol{\xi}(t) =\frac{d\mathbf{N}(t)}{dt},
\end{equation}
where $\mathbf{N}(t)$ is a three-dimensional Wiener process with a diffusion constant $\nu_{1}$.

Now, based on Assumption \ref{random force assumption 50-10}, the mathematically
rigorous form of Eq.(\ref{Langevin equation 50-40}) is the following
stochastic differential equations  \cite{Kallenberg1997}
\begin{equation}\label{Langevin equation 50-50}
\left\{
\begin{array}{ll}
d\mathbf{x}(t) = \mathbf{v}_{p}(t)dt, \\
md\mathbf{v}_{p}(t) =
-f\mathbf{v}_{p}(t)dt +
\mathbf{F}(\mathbf{x}, t)dt +
d\mathbf{N}(t),\\
\mathbf{x}(0)=\mathbf{x}_0, \quad  \mathbf{v}_{p}(0)=\mathbf{v}_0.
\end{array}
\right.
\end{equation}

\begin{assumption_my}\label{force assumption 3-10}
We assume that the functions
$\mathbf{F}(\mathbf{x}, t): R^3 \times R_{+} \rightarrow R^3$
 satisfy a global Lipschitz condition, i.e., for some constant $C_{0}$,
 \begin{math}|\mathbf{F}(\mathbf{x}_{1}, t) - \mathbf{F}(\mathbf{x}_{2}, t)|
\leq C_{0}|\mathbf{x}_{1} - \mathbf{x}_{2}|,\end{math}
for all $\mathbf{x}_{1}$ and $\mathbf{x}_{2}$ in $R^3$.
\end{assumption_my}

The following theorem is the main result of this section.
\begin{theorem_my}\label{convergence 50-10}
Suppose that Assumption \ref{random force assumption 50-10} and Assumption \ref{force assumption 3-10} are valid.
Then, at a time scale of an observer very large compare to the relaxation time $m/f$, the solution $\mathbf{x}(t)$ of the Langevin equation Eq.(\ref{Langevin equation 50-50})
converges to the solution $\mathbf{y}(t)$ of the Smoluchowski equation Eq.(\ref{Smoluchowski equation 50-10}) with probability one uniformly for t in compact subintervals of $[0, \infty)$ for all $\mathbf{v}_0$, i.e.,
\begin{equation}\label{convergence 50-20}
\lim_{m/f \rightarrow \infty} \mathbf{x}(t) = \mathbf{y}(t),
\end{equation}
where $\mathbf{y}(t)$ is the solution of the following Smoluchowski equation
\begin{equation}\label{Smoluchowski equation 50-10}
\begin{cases}
& \displaystyle d\mathbf{y}(t)
= \mathbf{b}(\mathbf{y}, t)dt + d\mathbf{w}(t), \\
& \mathbf{y}(0)=\mathbf{x}_0,
\end{cases}
\end{equation}
where $\mathbf{x}_0=\mathbf{x}(0)$, $\mathbf{w}(t)$ is a three-dimensional Wiener process with a diffusion coefficient $\nu_{0}$ determined by
\begin{equation}\label{diffusion 50-10}
\nu_{0} = \frac{\hbar_{0}}{2m},
\end{equation}
and \begin{math}\hbar_{0} = 2 m_{0} k_{0}T_{0}/(\rho_{0}q_{0}+f_{2}m_{0}).\end{math}
\end{theorem_my}

The proof of Theorem \ref{convergence 50-10} can be found in the appendix.
If $\hbar_{0} = \hbar$, then, we notice that Eq.(\ref{diffusion 50-10}) coincides with the hypothesis $\nu = \hbar/2m$ in Nelson's stochastic mechanics \cite{Nelson1966}.

\section{A generalized Hamilton-Jacobi equation}
Following Nelson \cite{Nelson1966}, we define the mean forward derivative $D\mathbf{y}(t)$
and the mean backward derivative $D_{*}\mathbf{y}(t)$. We also have another Smoluchowski equation as \cite{Nelson1966}
\begin{equation}\label{Smoluchowski equation 2-30}
 d\mathbf{y}(t) = \mathbf{b}_{*}(\mathbf{y}, t)dt + d\mathbf{w}_{*}(t),
\end{equation}
where $\mathbf{w}_{*}(t)$ has the same properties
as $\mathbf{w}(t)$ except that the
$d\mathbf{w}_{*}(t)$ are independent of the
$\mathbf{y}(s)$ with $s\geq t$.

Following Nelson \cite{Nelson1966}, we introduce the the following definitions of current velocity
$\mathbf{v}(t)$ and osmotic velocity $\mathbf{u}(t)$.
\begin{equation}\label{velocity 20-11}
 \mathbf{v} = \frac{1}{2}(\mathbf{b}+\mathbf{b}_{*}),\quad \mathbf{u} = \frac{1}{2}(\mathbf{b}-\mathbf{b}_{*}).
\end{equation}

We have the following result \cite{Nelson1966}:
\begin{equation}\label{osmotic velocity 20-20}
\mathbf{u} = \nu_{0} \frac{\nabla \rho}{\rho} = \nu_{0} \nabla(\ln\rho).
\end{equation}

From Eq.(\ref{osmotic velocity 20-20}), we introduce the following definition of osmotic potential $R_{1}$
\begin{equation}\label{osmotic velocity 80-420}
m\mathbf{u} = \nabla R_{1},
\end{equation}
where the osmotic potential $R_{1}$ is defined by
\begin{math}R_{1} \triangleq m \nu_{0} \ln \rho.\end{math}

Following Nelson\cite{Nelson1966}, we introduce the definition of the mean second
derivative $\mathbf{a}(t)$ of the stochastic process
$\mathbf{y}(t)$ as
\begin{equation}\label{mean second derivative 2-10}
 \mathbf{a}(t) = \frac{1}{2}DD_{*}\mathbf{y}(t) +  \frac{1}{2}D_{*}D\mathbf{y}(t).
\end{equation}

Similar to the deterministic Newtonian mechanics,
we can also introduce the following concept
of deterministic momentum field $\mathbf{p}_{d}(\mathbf{x},t)$ and  stochastic momentum field $\mathbf{p}_{s}(\mathbf{x},t)$ of the Brownian particle:
\begin{equation}\label{momentum 20-11}
\mathbf{p}_{d}(\mathbf{x},t) = m(t)\mathbf{v}(\mathbf{x},t), \mathbf{p}_{s}(\mathbf{x},t) = m(t)\mathbf{u}(\mathbf{x},t).
\end{equation}

From Eq.(\ref{monad 20-30}), we see that the mass $m(t)$ of the particle is increasing linearly. Thus, our results  departs from the Nelson's stochastic mechanics. We have the following result.
\begin{proposition_my}\label{momentum pdps 40-10}
If there exists a functions $S_{1}(\mathbf{x},t)$ such that
\begin{equation}\label{assumption grad 20-20}
\mathbf{p}_{d} = \nabla S_{1},
\end{equation}
then, the deterministic momentum field $\mathbf{p}_{d}(\mathbf{x},t)$ and  stochastic momentum field $\mathbf{p}_{s}(\mathbf{x},t)$ of the particle satisfy the following equations
\begin{eqnarray}
&& \frac{\partial \mathbf{p}_{d}(t)}{\partial t} =
\omega_{0}\mathbf{p}_{d} + \mathbf{F}  - \frac{1}{2m} \nabla (\mathbf{p}_{d}^{2}) \nonumber\\
&& \ \ \  \ \ \ \ \ \ \ \ \ \ + \frac{1}{2m} \nabla (\mathbf{p}_{s}^{2})
 + \nu_{0} \nabla^2 \mathbf{p}_{s}, \label{momentum pd 20-11}\\
&&  \frac{\partial \mathbf{p}_{s}(t)}{\partial t} = \omega_{0}\mathbf{p}_{s} + - \nu_{0}
\nabla^2 \mathbf{p}_{d}
- \frac{1}{m}\nabla(\mathbf{p}_{d}
\cdot \mathbf{p}_{s}), \label{momentum ps 20-12}
\end{eqnarray}
where $\omega_{0} \triangleq  \rho_{0} q_{0}/m_{0}(t)$.
\end{proposition_my}
{\bfseries{Proof of Proposition \ref{momentum pdps 40-10}}}.
According to Eq.(\ref{momentum 20-11}), we have the following relationships
\begin{eqnarray}
&& \frac{\partial \mbox{\upshape\bfseries{p}}_{d}(t)}{\partial t} = \frac{d m(t)}{d t}\mbox{\upshape\bfseries{v}} + m(t)\frac{\partial \mbox{\upshape\bfseries{v}}}{d t}, \label{momentum pd 20-21}\\
&& \frac{\partial \mbox{\upshape\bfseries{p}}_{s}(t)}{\partial t} = \frac{d m(t)}{d t}\mbox{\upshape\bfseries{u}} + m(t)\frac{\partial \mbox{\upshape\bfseries{u}}}{d t}. \label{momentum pd 20-22}
\end{eqnarray}

Following a similar method of Nelson\cite{Nelson1966}, we obtain the following results
\begin{eqnarray}
&&\frac{\partial \mathbf{v}}{\partial t} =
\frac{\mathbf{F}}{m} - (\mathbf{v}
\cdot \nabla)\mathbf{v} +
(\mathbf{u} \cdot
\nabla)\mathbf{u} + \nu_{0} \nabla^2
\mathbf{u}, \label{Nelson equation 20-21}\\
&&\frac{\partial \mathbf{u}}{\partial t} = - \nu_{0}
\nabla(\nabla \cdot \mathbf{v})
- \nabla(\mathbf{v}
\cdot \mathbf{u}). \label{Nelson equation 20-22}
\end{eqnarray}

Putting Eq.(\ref{Nelson equation 20-21}-\ref{Nelson equation 20-22}) into Eq.(\ref{momentum pd 20-21}-\ref{momentum pd 20-22}), we obtain
\begin{equation}\label{momentum pd 20-31}
\frac{\partial \mbox{\upshape\bfseries{p}}_{d}(t)}{\partial t} = \frac{d m(t)}{d t}\mbox{\upshape\bfseries{v}} + \mbox{\upshape\bfseries{F}}
- m(\mbox{\upshape\bfseries{v}}\cdot \nabla)\mbox{\upshape\bfseries{v}}
+ m(\mbox{\upshape\bfseries{u}} \cdot
\nabla)\mbox{\upshape\bfseries{u}} + m \nu_{0} \nabla^2
\mbox{\upshape\bfseries{u}},
\end{equation}
\begin{equation}\label{momentum pd 20-32}
\frac{\partial \mbox{\upshape\bfseries{p}}_{s}(t)}{\partial t} = \frac{d m(t)}{d t}\mbox{\upshape\bfseries{u}}
- m \nu_{0}\nabla(\nabla \cdot \mbox{\upshape\bfseries{v}})
- m \nabla(\mbox{\upshape\bfseries{v}} \cdot \mbox{\upshape\bfseries{u}}).
\end{equation}

Using Eq.(\ref{monad 20-30}) and some formula in field theory, we arrive at Eq.(\ref{momentum pd 20-11}-\ref{momentum ps 20-12}).
This ends the proof of Proposition \ref{momentum pdps 40-10}. $\square$

We may call the functions $S_{1}(\mathbf{x},t)$ defined in Eq.(\ref{assumption grad 20-20}) as the current potential.
The current potential $S_{1}(\mathbf{x},t)$ is not uniquely defined by the deterministic momentum field $\mathbf{p}_{d}(\mathbf{x},t)$.

\begin{theorem_my}\label{generalized Hamilton-Jacobi 40-10}
Suppose that there exist two functions $V(\mathbf{x})$ and $S_{1}$ such that
\begin{equation}\label{assumption grad 20-15}
\mathbf{F}(\mathbf{x},t)= - \nabla V(\mathbf{x}),
\end{equation}
\begin{equation}\label{assumption grad 20-25}
\mathbf{p}_{d} = \nabla S_{1}.
\end{equation}
Then, the generalized Hamilton's principal function $S \triangleq S_{1} -i R_{1}$, $i^{2} = -1$,
satisfies the following generalized Hamilton-Jacobi equation
\begin{equation}\label{ghj 20-10}
-\frac{\partial S}{\partial t} = -\omega_{0}S + \frac{1}{2m} (\nabla S)^2
+ V(\mathbf{x}) - i \nu_{0}\nabla^{2} S  + a_{1}(t) + i a_{2}(t),
\end{equation}
where $a_{1}(t)$ and $a_{2}(t)$ are two unknown real functions of $t$.
\end{theorem_my}

The proof of Theorem \ref{generalized Hamilton-Jacobi 40-10} can be found in the appendix. The generalized Hamilton's principal function $S$ is not uniquely defined by $\mathbf{p}_{d}$. The reason is that $\mathbf{p}_{d} = \nabla S_{1}$. Thus, $S_{1}$ is not uniquely defined by $\mathbf{p}_{d}$.

It is not surprising that the generalized Hamilton-Jacobi equation Eq.(\ref{ghj 20-10}) is similar to the following Hamilton-Jacobi equation in classical mechanics \cite{ZengJY2007}
\begin{equation}\label{Hamilton-Jacobi 10-10}
-\frac{\partial S}{\partial t} = \frac{1}{2m} (\nabla S)^2  + V(\mathbf{x}).
\end{equation}

Similar to Bohr's Correspondence Principle in quantum mechanics, we may also introduce the following correspondence principle in stochastic mechanics.
\begin{assumption_my}\label{correspondence principle 20-10}
If the diffusion constant $\nu_{0}$ and the parameter $\omega_{0}$ are small enough, i.e., $\nu_{0} \rightarrow 0$ and $\omega_{0} \rightarrow 0$, then, the generalized Hamilton-Jacobi equation Eq.(\ref{ghj 20-10}) in stochastic mechanics becomes identical to the Hamilton-Jacobi equation Eq.(\ref{Hamilton-Jacobi 10-10}) in classical mechanics
\end{assumption_my}

\begin{theorem_my}\label{theorem ghj 20-20}
Suppose that Eq.(\ref{assumption grad 20-15}-\ref{assumption grad 20-25}) are valid.
Then, the generalized Hamilton's principal function $S$
satisfies the following generalized Hamilton-Jacobi equation
\begin{equation}\label{generalized Hamilton-Jacobi 20-50}
-\frac{\partial S}{\partial t} = -\omega_{0}S + \frac{1}{2m} (\nabla S)^2
+ V(\mathbf{x}) - i \nu_{0}\nabla^{2} S.
\end{equation}
\end{theorem_my}
{\bfseries{Proof of Theorem \ref{theorem ghj 20-20}}}
Let $\omega_{0}=\nu_{0} = 0$. Then, from Eq.(\ref{osmotic velocity 20-20}), we have $\mathbf{u} = 0$. Thus, from Eq.(\ref{momentum 20-11}), we have $\mathbf{p}_{s} = 0$. Then, from Eq.(\ref{osmotic velocity 80-420}),  $R_{1}$ is a constant. Thus, Eq.(\ref{ghj 20-10}) can be written as
\begin{equation}\label{hj 20-10}
-\frac{\partial S_{1}}{\partial t} = \frac{1}{2m} (\nabla S_{1})^2 + V(\mathbf{x}) + a_{1}(t) + i a_{2}(t),
\end{equation}

According to Assumption \ref{correspondence principle 20-10}, Eq.(\ref{hj 20-10}) should be identical to the Hamilton-Jacobi equation Eq.(\ref{Hamilton-Jacobi 10-10}). Thus, we obtain $a_{1}(t) = 0$
and $a_{2}(t) = 0$. This ends the proof of Theorem \ref{theorem ghj 20-20}. $\square$

%\subsection{Subsection 1}
%\subsection{Subsection 2}
\section{Generalized Schr\"{o}dinger equation in stochastic Newtonian mechanics}
We introduce the following definition
\begin{equation}\label{wave function 30-10}
\psi(\mathbf{x},t) = e^{\frac{i S(\mathbf{x},t)}{2m\nu_{0}}}.
\end{equation}

We may call the function $\psi(\mathbf{x},t)$ as wave function following Hamiltonian mechanics\cite{Goldstein2002}. The generalized Hamilton's principal function $S$ is not uniquely defined by $\mathbf{p}$. Therefore, the wave function $\psi(\mathbf{x},t)$ defined by Eq.(\ref{wave function 30-10}) is not uniquely defined by $\mathbf{p}$.

\begin{theorem_my}\label{Schrodinger like 30-10}
The wave function $\psi(\mathbf{x},t)$
defined by Eq.(\ref{wave function 30-10}) satisfies the following Schr\"{o}dinger like equation
\begin{equation}\label{Schrodinger 30-10}
i  \frac{\partial \psi}{\partial t} = \omega_{0} \psi \ln \psi - \nu_{0} \nabla^2\psi
+ \frac{1}{2m\nu_{0}} V\psi.
\end{equation}
Eq.(\ref{Schrodinger 30-10}) is equivalent to the generalized Hamilton-Jacobi equation Eq.(\ref{generalized Hamilton-Jacobi 20-50}).
\end{theorem_my}
{\bfseries{Proof of Theorem \ref{Schrodinger like 30-10}}}.
From the definition Eq.(\ref{wave function 30-10}), we have
\begin{equation}\label{Hamilton function 30-10}
S(\mathbf{x},t) = \frac{2m\nu_{0}}{i} \ln \psi(\mathbf{x},t).
\end{equation}

Putting Eq.(\ref{Hamilton function 30-10}) into Eq.(\ref{generalized Hamilton-Jacobi 20-50}),
we obtain a Schr\"{o}dinger like equation Eq.(\ref{Schrodinger 30-10}).
Conversely, putting Eq.(\ref{wave function 30-10})
into Eq.(\ref{Schrodinger 30-10}), we obtain the generalized Hamilton-Jacobi equation Eq.(\ref{generalized Hamilton-Jacobi 20-50}). This ends the proof of Theorem \ref{Schrodinger like 30-10}. $\square$

Putting Eq.(\ref{diffusion 50-10}) into Eq.(\ref{Schrodinger like 30-10}), we obtain the following proposition.
\begin{wcorollary_my}\label{reduce Schrodinger 30-20}
Suppose that Eq.(\ref{assumption grad 20-15}-\ref{assumption grad 20-25}) are valid.
Then, the wave function $\psi(\mathbf{x},t)$
defined by Eq.(\ref{wave function 30-10}) satisfies the following nonlinear Schr\"{o}dinger equation
\begin{equation}\label{Schrodinger 50-10}
i \hbar_{0} \frac{\partial \psi}{\partial t} =
-\frac{\hbar_{0}^2}{2m}\nabla^2\psi + V\psi + \hbar_{0}\omega_{0} \psi \ln \psi.
\end{equation}
\end{wcorollary_my}

The mass-increasing effect indicated in Eq.(\ref{monad 20-30}) is so small that it may be difficult for us to detect in the time scale of human beings. If we introduce the following assumption: $\omega_{0} \rightarrow 0$, then, the nonlinear Schr\"{o}dinger equation Eq.(\ref{Schrodinger 50-10}) reduces to the Schr\"{o}dinger equation Eq.(\ref{Schrodinger equation 10-10}).

\section{Discussion}
The wave function $\psi(\mathbf{x},t)$ is not uniquely defined by $\mathbf{p}$. Wallstrom \cite{WallstromTC1994} points out:"the Madelung equation are not equivalent to the Schr\"{o}dinger equation unless a quantization imposed. This condition is that the wave function be single valued;" This quantization condition was confirmed by plenty of experiments in quantum mechanics\cite{Landau-Lifshitz1958,ZengJY2007}. Thus, a successful stochastic interpretation of nonrelativistic quantum phenomena should derive this quantization condition.

\section{Conclusion}
In Nelson's stochastic mechanics, the Schr\"{o}dinger equation
Eq.(\ref{Schrodinger equation 10-10}) is derived based on two main hypotheses. In previous works of stochastic mechanics, these two hypotheses can not be explained. Based on a fluidic continuum model of the vacuum and a sink flow model of microscopic particles, we show that Nelson's two hypotheses can be derived. Similar to Bohr's Correspondence Principle, we also introduce a correspondence principle. The generalized Hamilton's principal function satisfies a generalized Hamilton-Jacobi equation. Further, considering the mass-increasing effects, we show that there exists a nonlinear Schr\"{o}dinger equation for microscopic particles. As a byproduct, the Planck constant $h$ is calculated theoretically.

\section*{Acknowledgments}
This work was supported by the Doctor Research Foundation of Henan Polytechnic University (Grant No. 648734).

\section{Appendix}
\subsection{Proof of Theorem \ref{convergence 50-10}}
Using the following definitions
\begin{equation}\label{three definitions 50-10}
 \beta = \frac{f}{m}, \quad \mathbf{K}(\mathbf{x}, t)
 =  \frac{\mathbf{F}(\mathbf{x}, t)}{m},\quad \mathbf{B}(t) = \frac{\mathbf{N}(t)}{m},
\end{equation}
we see that $\mathbf{B}(t)$ is a three-dimensional
Wiener process with a diffusion coefficient  \cite{Kallenberg1997}
\begin{equation}\label{diffusion coefficient 3-42}
  \nu_{2} =  \frac{\nu_{1}}{m^2} = \frac{f k_{0}T_{0}}{m^2} = \frac{\beta k_{0}T_{0}}{m}.
\end{equation}

Using Eq.(\ref{three definitions 50-10}), Eq.(\ref{Langevin equation 50-50}) become
\begin{equation}\label{Langevin equation 50-21}
\left\{
\begin{array}{ll}
\displaystyle   d\mathbf{x}(t) = \mathbf{v}_{p}(t)dt, \\
d\mathbf{v}_{p}(t)
 = -\beta \mathbf{v}_{p}(t)dt
 +\mathbf{K}(\mathbf{x}, t)dt
 + d\mathbf{B}(t).
\end{array}
\right.
\end{equation}

We define
\begin{equation}\label{three definitions 50-30}
 \mathbf{b}(\mathbf{x}, t)
 =  \frac{\mathbf{K}(\mathbf{x}, t)}{\beta},
 \quad \mathbf{w}(t) = \frac{\mathbf{B}(t)}{\beta}.
\end{equation}
Then, we see that $\mathbf{w}(t)$ is a three-dimensional
Wiener process with a diffusion coefficient  \cite{Kallenberg1997}
\begin{equation}\label{diffusion coefficient 50-40}
  \nu_{0} =  \frac{\nu_{2}}{\beta^2} = \frac{ k_{0} T_{0}}{m \beta} = \frac{ k_{0}
  T_{0}}{f}.
\end{equation}

Using Eq.(\ref{three definitions 50-30}),  Eq.(\ref{Langevin equation 50-21}) can be written as
\begin{equation}\label{Langevin equation 50-31}
\left\{
\begin{array}{ll}
\displaystyle d\mathbf{x}(t) = \mathbf{v}_{p}(t)dt, \\
d\mathbf{v}_{p}(t) = -\beta \mathbf{v}_{p}(t)dt
 + \beta \mathbf{b}(\mathbf{x}, t)dt
 + \beta d\mathbf{w}(t).
\end{array}
\right.
\end{equation}

 Let $\mathbf{x}(t)$ be the solution of Eq.(\ref{Langevin equation 50-31}) with $\mathbf{x}(0)=\mathbf{x}_0,
 \mathbf{v}_{p}(0)=\mathbf{v}_0$.
According to Assumption \ref{force assumption 3-10},
the functions $\mathbf{b}(\mathbf{x}, t): R^3 \times R_{+} \rightarrow R^3$ also satisfies a global Lipschitz condition. Applying Nelson's Theorem 10.1 in  \cite{Nelson1972},
for a time scale of an observer very large compared to the relaxation time $1/\beta$,
$\mathbf{x}(t)$ converges to the solution
$\mathbf{y}(t)$ of the Smoluchowski equation
\begin{equation}\label{Smoluchowski equation 3-20}
 d\mathbf{y}(t) = \mathbf{b}(\mathbf{y}, t)dt + d\mathbf{w}(t)
\end{equation}
with $\mathbf{y}(0)=\mathbf{x}_0$.

From Eq.(\ref{Langevin equation 50-40}) and Eq.(\ref{diffusion coefficient 50-40}), we have
\begin{equation}\label{fhfdh}
f = \left ( \frac{ \rho_{0} q_{0}}{m_{0}} + f_{2} \right ) m
\end{equation}
\begin{equation}\label{diffusion coefficient 50-50}
\nu_{0} = \frac{k_{0}T_{0}}{f} = \frac{m_{0}k_{0}T_{0}}{(\rho_{0}q_{0}+f_{2}m_{0})m}.
\end{equation}

We introduce the following notations
\begin{equation}\label{Planck constant definition 50-10}
\hbar_{0} = \frac{2 m_{0} k_{0}T_{0}}{(\rho_{0}q_{0}+f_{2}m_{0})}= \frac{h_{0}}{2\pi}.
\end{equation}

Using Eq.(\ref{Planck constant definition 50-10}), Eq.(\ref{diffusion coefficient 50-50}) can be written as
\begin{equation}\label{diffusion coefficient and Planck constant 3-10}
  \nu_{0} = \frac{\hbar_{0}}{2m}, \quad or, \ \nu_{0} = \frac{h_{0}}{4 \pi m}.
\end{equation}
This ends the proof of Theorem \ref{convergence 50-10}. $\square$

\subsection{Proof of Theorem \ref{generalized Hamilton-Jacobi 40-10}}
Multiplying Eq.(\ref{momentum pd 20-11}) with $-1$
and adding Eq.(\ref{momentum ps 20-12}) multiplied by $i$, we obtain
\begin{eqnarray}\label{ghj 20-20}
&& -\frac{\partial (\mbox{\upshape\bfseries{p}}_{d} -i \mbox{\upshape\bfseries{p}}_{s})}{\partial t}
  =  - \omega_{0} (\mbox{\upshape\bfseries{p}}_{d} -i \mbox{\upshape\bfseries{p}}_{s}) - \mbox{\upshape\bfseries{F}} \nonumber \\
&& + \frac{1}{2m} \nabla [(\mathbf{p}_{d} -i \mathbf{p}_{s})^2] - i \nu_{0}\nabla^{2} (\mbox{\upshape\bfseries{p}}_{d} -i \mbox{\upshape\bfseries{p}}_{s}).
\end{eqnarray}

We introduce the following definition
\begin{equation}\label{momentum p 20-10}
\mathbf{p} \triangleq \mathbf{p}_{d}
-i \mathbf{p}_{s}.
\end{equation}

Putting Eq.(\ref{momentum p 20-10}) into Eq.(\ref{ghj 20-20}), we have
\begin{equation}\label{ghj 20-30}
-\frac{\partial \mathbf{p}}{\partial t}
= - \omega_{0}\mbox{\upshape\bfseries{p}} - \mathbf{F} + \frac{1}{2m} \nabla (\mathbf{p}^2) - i \nu_{0}\nabla^{2} \mathbf{p}.
\end{equation}

We introduce the following definition
\begin{equation}\label{Hamilton function s 20-10}
S \triangleq S_{1} -i R_{1}.
\end{equation}

Putting Eq.(\ref{osmotic velocity 80-420}) and  Eq.(\ref{assumption grad 20-25}) into Eq.(\ref{momentum p 20-10}) and using Eq.(\ref{Hamilton function s 20-10}), we have
\begin{equation}\label{nabla 20-10}
\mathbf{p} = \nabla S.
\end{equation}

Putting Eq.(\ref{nabla 20-10}) into Eq.(\ref{ghj 20-30}), we obtain
\begin{equation}\label{ghj 20-40}
-\frac{\partial (\nabla S)}{\partial t} =  - \omega_{0}\nabla S - \mathbf{F} + \frac{1}{2m} \nabla [(\nabla S)^2]   - i \nu_{0}\nabla^{2} (\nabla S).
\end{equation}

Noticing $\mathbf{F} = - \nabla V(\mathbf{x})$, Eq.(\ref{ghj 20-40}) becomes
\begin{equation}\label{ghj 20-50}
-\frac{\partial (\nabla S)}{\partial t}
=  - \omega_{0}\nabla S + \nabla V + \frac{1}{2m} \nabla[(\nabla S)^2]   - i \nu_{0}\nabla^{2} (\nabla S).
\end{equation}

Eq.(\ref{ghj 20-50}) can be written as
\begin{equation}\label{ghj 20-60}
\nabla \left [\frac{\partial S}{\partial t} - \omega_{0} S
+  V(\mathbf{x}) + \frac{1}{2m} (\nabla S)^2   - i \nu_{0}\nabla^{2} S\right ] = 0.
\end{equation}

Integration of Eq.(\ref{ghj 20-60}) gives
\begin{equation}\label{ghj 20-70}
- \frac{\partial S}{\partial t}
= - \omega_{0} S +  V(\mathbf{x}) + \frac{1}{2m} (\nabla S)^2   - i \nu_{0}\nabla^{2} S + a_{1}(t) + i a_{2}(t),
\end{equation}
where  $a_{1}(t)$ and  $a_{2}(t)$ are two unknown real functions of $t$.
This ends the proof of Theorem \ref{generalized Hamilton-Jacobi 40-10}. $\square$

\providecommand{\noopsort}[1]{}\providecommand{\singleletter}[1]{#1}%

%\end{CJK*}
\end{document}